\documentclass{emulateapj}
\usepackage{graphicx} 
\usepackage{epstopdf}
\usepackage{longtable}

\shorttitle{PG~1553+113: five years of observations with MAGIC}
\shortauthors{Aleksi\'c et al.}

\begin{document}


\title{PG 1553+113: FIVE YEARS OF OBSERVATIONS with MAGIC}


% authors 1.7.2011  Format IoP-ApJ
%
\author{
J.~Aleksi\'c$^{1}$,
E.~A.~Alvarez$^{2}$,
L.~A.~Antonelli$^{3}$,
P.~Antoranz$^{4}$,
M.~Asensio$^{2}$,
M.~Backes$^{5}$,
J.~A.~Barrio$^{2}$,
D.~Bastieri$^{6}$,
J.~Becerra Gonz\'alez$^{7,8}$,
W.~Bednarek$^{9}$
% ...
F.~D'Ammando$^{34}$
}
\address{$^{1}$ IFAE, Edifici Cn., Campus UAB, E-08193 Bellaterra, Spain}
\address{$^{2}$ Universidad Complutense, E-28040 Madrid, Spain}
\address{$^{3}$ INAF National Institute for Astrophysics, I-00136 Rome, Italy}
\address{$^{4}$ Universit\`a  di Siena, and INFN Pisa, I-53100 Siena, Italy}
\address{$^{5}$ Technische Universit\"at Dortmund, D-44221 Dortmund, Germany}
\address{$^{6}$ Universit\`a di Padova and INFN, I-35131 Padova, Italy}
\address{$^{7}$ Inst. de Astrof\'{\i}sica de Canarias, E-38200 La Laguna, Tenerife, Spain}
\address{$^{8}$ Depto. de Astrof\'{\i}sica, Universidad de La Laguna, E-38206 La Laguna, Spain}
\address{$^{9}$ University of \L\'od\'z, PL-90236 Lodz, Poland}
% ...
\address{$^{34}$ Consorzio Interuniversitario Fisica Spaziale (CIFS), v.le Settimio Severo 63, I-10133 Torino, Italy}
% ...
                   
\begin{abstract}
We present the results of five years (2005--2009) of MAGIC observations of the BL Lac object PG~1553+113 at very high energies (VHEs, E $>$ 100\,GeV). 
Power law fits of the individual years are compatible with a steady mean photon index $\Gamma$~=~4.27~$\pm$~0.14. 
In the last three years of data, the flux level above 150\,GeV shows a clear variability (probability of constant flux $<$ 0.001\%). The flux variations are modest, lying in the range from 4\% to 11\% of the Crab Nebula flux. Simultaneous optical data also show only modest variability that seems to be correlated with VHE gamma ray variability. We also performed a temporal analysis of (all available) simultaneous {\it Fermi}/LAT data of PG~1553+113 above 1\,GeV, which reveals hints of variability in the 2008--2009 sample. 
Finally, we present a combination of the mean spectrum measured at very high energies with archival data available for other wavelengths. The mean spectral energy distribution can be modeled with a one--zone Synchrotron Self Compton (SSC) model, which gives the main physical parameters governing the VHE emission in the blazar jet.
\end{abstract}


\section{Introduction}
The majority of extragalactic $\gamma$ ray sources, both at GeV energies and above 100\,GeV, are blazars, radio-loud active galactic nuclei with a relativistic jet pointing towards the Earth.
Their emission is dominated by the non-thermal continuum produced within the jet and boosted by relativistic effects \citep{urry1995}. The spectral energy distribution (SED) displays two broad peaks, widely interpreted as due to synchrotron, low frequency peak, and inverse Compton, high frequency peak, mechanism (although the high energy peak could also be the result of hadronic processes, as proposed in \citealt{mannheim93}). Among blazars, BL Lac objects are characterized by extremely weak emission lines in their optical spectra, which often makes a measurement of their redshift difficult. The large majority of extragalactic sources detected above 100\,GeV are BL Lac objects, in which the peak of the synchrotron bump is located in the UV-X-ray bands and the high energy peak around 100\,GeV (these sources are often called high frequency peaked BL Lacs, HBLs). 
PG~1553+113 is a BL Lac discovered by \citet{green86}. The large X-ray to radio flux ratio makes this source a typical HBL. Indeed, its synchrotron peak is located between the UV and X-ray bands. Its optical spectrum is featureless, preventing the direct determination of the redshift. Indirect methods based on the non detection of the characteristic lines and of the 
host galaxy provide lower limits, ranging from 0.09 to 0.78 \citep{sbarufatti2006,sbarufatti2005}. 
The most recent estimate, based on the Ly alpha forest method, gives a $z~\sim$~0.40--0.45 \citep{danforth10}.

PG~1553+133 has been discovered as a VHE $\gamma$ ray emitter by H.E.S.S. \citep{aharonian06a} and MAGIC \citep{albert07}, with a flux of approximately 2\% of that of the Crab Nebula above 200\,GeV. The spectrum appears extremely soft (photon index $\Gamma\sim$\,4), as expected by the absorption of VHE photons through interaction with the Extragalactic Background Light (EBL) if the source is located at relatively large redshift \citep{stecker92}. 
The absorption process, in fact, is a function of the energy of the photon and of the distance it has traveled. Spectra with indices $\Gamma\sim$\,4 have been observed in blazars located at redshifts above 0.2.

VHE $\gamma$ ray observations have been used as alternative method to constrain the distance of blazars.  \citet{aharonian06b} proposed a way to set an upper limit on the distance of blazars based on the assumption that the VHE intrinsic spectrum, obtained by correcting the observed spectrum for the extragalactic background light absorption, cannot be harder than a fixed value given by theory. The technique, applied to PG~1553+113, lead to an upper limit of $z~<$~0.74.
Recently, \citet{prandini10} extended this method using the spectrum measured at lower energies as limiting slope for the original spectrum, obtaining $z~<$~0.66 for PG~1553+113, at 2\,$\sigma$ level. Other approaches require the absence of a pile up at high energies. With this method, \citet{mazin07} get $z~<$~0.42. 
Hence, in the case of PG~1553+113, the upper limits obtained with these methods are in the range of the limits set by optical measurements. In this work we adopt the redshift $z$~=~0.40. Such a large redshift is also supported by  the absence of significant points at energies above 700\,GeV in the spectrum of the source. 

At MeV-GeV energies, PG~1553+113 was not detected by EGRET, but it is well visible by the Large Area Telescope (LAT) on-board {\it Fermi}, being detected with a significance above 10\,$\sigma$ already in the first three months of observations \citep{abdo09}. \citet{abdo10b} show that the {\it Fermi}/LAT spectrum is surprisingly constant both in normalization and slope over $\sim$\,200 days. Interestingly, a stability of the spectrum was also suggested by the H.E.S.S. and MAGIC observations, showing rather marginal variability during 2005 and 2006 observations. The stability of the VHE $\gamma$ ray emission is in contrast to the behavior commonly observed in other TeV emitting BL Lacs, showing rather pronounced variations at all timescales.

After its discovery, PG~1553+113 was regularly observed by MAGIC. In this paper, we present  the analysis of the new data taken from 2007 to 2009, combined with previous observations. The differential and integral fluxes are analyzed, in comparison with partially simultaneous  measurements at other wavelengths, and the stability of the spectrum over this long period is studied. Finally, we combine all the data available and model the SED with a one--zone Synchrotron Self Compton (SSC) model, constraining the main physical parameters that govern the VHE emission in the blazar jet.

\section{MAGIC Observations and Data Analysis}
Since autumn 2009, MAGIC \citep{cortina09} is a stereo system composed by two 
Imaging Atmospheric Cherenkov Telescopes (IACTs) located on La Palma,
Canary Islands, Spain ($28.75^{\circ}$N, $17.89^{\circ}$W, 2240\,m~asl).
In this paper, we present only data collected before the stereo
upgrade, with a single telescope, MAGIC~I \citep{baixeras04}, 
hereafter called MAGIC. 
The parabolic-shaped reflector,
with a total mirror area of 236\,m$^{2}$, allows MAGIC to collect 
the Cherenkov light and focus it onto a multi-pixel
camera, composed of 577 photo-multipliers. MAGIC 
camera and trigger are designed to record data not only during dark nights, but
also under moderate light conditions (i.e. moderate moon, twilight).
Due to its comparatively low trigger 
energy threshold of $\sim$\,50\,GeV, 
MAGIC is well suited to perform multiwavelength observations together with instruments
operating in the GeV range.

\begin{table}[tbh]
  \centering
  \caption{PG~1553+113 FINAL DATA SET}
  \begin{tabular}{cccccc}
  \hline
  \hline
  Cycle   & Date           & Eff. Time & $Zd$         & Rate  &  DC         \\
          &                &   [min]   &[$^{\circ}$] & [Hz]  & [$\mu$A]    \\
  \hline 
  \hline
          &  23/03/2007    &  58       &  19 - 29  &   164  &  dark night \\
          &  19/04/2007    &  32       &  22 - 28  &   155  &  dark night \\
          &  20/04/2007    &  150      &  17 - 29  &   163  &  dark night \\
   III    &  21/04/2007    &  115      &  17 - 24  &   155  &  dark night \\
          &  22/04/2007    &  101      &  17 - 23  &   162  &  dark night \\
          &  23/04/2007    &  143      &  17 - 27  &   161  &  dark night \\
          &  24/04/2007    &  92       &  17 - 23  &   160  &  dark night \\
 \hline
          &  17/03/2008    &   58      &  17 - 19  & 150  &  dark night \\
          &  18/03/2008    &   26      &  18 - 19  & 150  &  dark night \\
          &  01/04/2008    &   43      &  20 - 26  & 167  &  dark night \\
          &  05/04/2008    &   109     &  17 - 31  & 167  &  dark night \\
  IV      &  13/04/2008    &   97      &  17 - 22  & 147  &  dark night \\
          &  29/04/2008    &   44      &  27 - 36  & 151  &  dark night \\
          &  03/05/2008    &   24      &  26 - 31  & 146  &  dark night \\
          &  04/05/2008    &   40      &  28 - 36  & 155  &  dark night \\
          &  05/05/2008    &   38      &  26 - 33  & 150  &  dark night \\
          &  07/05/2008    &   40      &  28 - 36  & 153  &  dark night \\
  \hline 
          &  16/04/2009    &  93       &  17 - 27 & 133  & 2.8 - 3.7 \\
          &  17/04/2009    &  103      &  17 - 28 & 151  & 1.6 - 2.4 \\
  V       &  18/04/2009    &  126      &  17 - 28 & 168  & 0.7 - 1.7 \\
          &  20/04/2009    &  73       &  19 - 35 & 171  & 0.9 - 1.4 \\
          &  21/04/2009    &  57       &  23 - 34 & 177  & 0.8 - 1.0 \\
          &  15/06/2009    &  57       &  24 - 35 & 125  & 0.8 - 2.1 \\  
\hline
\hline
 \end{tabular} 
  \footnotetext{PG~1553+113 data set from 2007 to 2009 used in this study. From left to right: MAGIC Cycle of observation, first column, and corresponding dates in dd/mm/yy, second column; effective time of observation in minutes and zenith angle range in degrees, third and fourth column. In the last two columns, the rate of the events after the image cleaning, in Hz, and the mean DC current in the camera, in unit of $\mu$A, are shown. The night is considered as dark night, if the DC current while observing an extragalactic object is less than indicatively 1.2 $\mu$A.}  \label{table_obs}
\vspace{0.16cm}
  \end{table}

The total Field of View (FoV) of the MAGIC camera is $3.5^{\circ}$, and the effective 
collection area is
of the order of $10^{5}$\,m$^{2}$ at 200\,GeV for a source close to zenith.
The incident light pulses are converted into analog signals,
transmitted via optical fibers and digitised by 2\,GHz fast analog to
digital converters (FADCs).

PG~1553+113 was observed with the MAGIC telescope for 
nearly 19~hours in 2005 and 2006  \citep{albert07}; it was also the subject of a 
multiwavelength campaign carried out in July 2006
with optical, X-ray and TeV $\gamma$ ray telescopes \citep{albert09}. 
Here, we present the results of follow-up observations, 
performed for 14~hours in March-April 2007, for nearly 26~hours in March-May 2008, 
some of those simultaneously with other 
instruments \citep{aleksic10}, and for about 24~hours in March-July 2009, which were partly
taken in moderate light conditions (moon light).
Unfortunately, both 2008 and 2009 observations were severely affected 
by bad weather (including \textit{calima}, i.e. 
Saharan sand-dust in the atmosphere)
that limited the final data set and resulted in an
increased energy threshold. 

All data analyzed here were taken in the 
false-source tracking (wobble) mode \citep{fomin94},
in which the telescope pointing was alternated every 20 minutes between
two sky positions at $0.4^\circ$ offset from the source.
The zenith angle of 2007 observations varied
from $17^\circ$ to $30^\circ$, in 2008
it extended up to $36^\circ$, while in 2009 it covered the 
range from $17^\circ$ to $35^\circ$. 

The data were analyzed  using the standard MAGIC 
analysis chain \citep{albert08a,aliu09}.
Severe quality cuts based on event rate after night sky background
suppression were applied to the sample; 28.7~hours of good quality
data remained after these cuts, out of which 11.5~hours were taken in 2007,
8.7~hours in 2008 and 8.5 hours in 2009. 
More details about the final data set can be found in Table~\ref{table_obs}.
For the signal study, a cut in the parameter 
\emph{size} removed  events with a total charge less than 80 
photo-electrons (phe) in the 2007 data set, and 200\,phe in 2008 and 2009
data sets. In the latter case, this cut reduces the
effect of the moon light.

Finally, for the spectrum determination an additional cut in PMT DC current,
namely above 2.5~$\mu$A, was applied to the 2009 sample, 
in order to reduce systematics due to
the moon light \citep{britzger09}, resulting in 6.9\,hours
of good quality data.
For the conclusive steps of the analysis, Monte Carlo (MC) simulations
of $\gamma$-like events were used. Hadronic background suppression was achieved
using the Random Forest (RF) method \citep{albert08c}, in which each event is assigned 
an additional parameter, the \emph{hadronness},
which is related to the probability that the event is not $\gamma$-like.
The RF method was also used in the energy estimation.
The threshold of the analysis was estimated to be 
80\,GeV in 2007, 150\,GeV in 2008 and 160\,GeV in 2009, 
as shown in Table~\ref{table_signal}.

\begin{table}[t!]
  \centering 
  \caption{PG~1553+113 SIGNAL}
  \begin{tabular}{c|c|c|c|c|c}
  \hline
 \hline
  Year& Time& Opt. PSF  &Energy Th. &Excesses & Signif.     \\
         & [h] &    [mm]      &[GeV] &         &  [$\sigma$] \\
 \hline 
 \hline
    2007 & 11.5 &     13     &80 & 1400 $\pm$ 242  &  5.8  \\
 \hline 
    2008 & 8.7 &      13     &150 &  542 $\pm$ 69  & 8.1 \\
 \hline 
    2009 & 8.5 &     14.8    &160  & 212 $\pm$ 52 & 4.2 \\
 \hline 
 \hline
   \end{tabular}
   \footnotetext{PG~1553+113 signal study. From left to right: year of observation, effective  time of good quality data used for  the signal analysis, optical point spread function (PSF), energy threshold of the analysis, number of excess events observed and significance of the signal.}  \label{table_signal}
\vspace{0.4cm}
  \end{table}

Due to changes in the telescope performance, the {\it sigma} of 
the optical point-spread function (PSF) of 2007 and 2008  
was measured to be 13.0\,mm, while in 2009 it was 14.9\,mm. 
To take all these differences into account, the 
data were analysed separately, using
dedicated sets of simulated data.

 \begin{table*}[htb!]
  \centering
  \caption{PG~1553+113 MEASURED SPECTRA}
  \begin{tabular}{c|c|c|c|c}
  \hline
 \hline
   Year  & $F$ $>$ 150\,GeV             &  $F$ $>$ 150\,GeV &   f$_0$   & $\Gamma$  \\
         & [ cm$^{-2}$ s$^{-1}$]  & [Crab \%]         &[ cm$^{-2}$ s$^{-1}$ TeV$^{-1}$]   &   \\
\hline 
 \hline
    2007 &  (1.40 $\pm$ 0.38) $\cdot$ 10$^{-11}$ & 4\%  & (1.1 $\pm$ 0.3) $\cdot$ 10$^{-10}$  & 4.1 $\pm$ 0.3 \\
 \hline 
    2008 &  (3.70 $\pm$ 0.47) $\cdot$ 10$^{-11}$  & 11\% & (2.6 $\pm$ 0.3) $\cdot$ 10$^{-10}$ & 4.3 $\pm$ 0.4 \\
 \hline 
    2009 & (1.63 $\pm$ 0.45) $\cdot$ 10$^{-11}$  &  5\% & (1.3 $\pm$ 0.2) $\cdot$ 10$^{-10}$ & 3.6 $\pm$ 0.5 \\
 \hline 
 \hline  
\end{tabular}  
 \tablecomments{Spectra of the individual years of observations of  PG~1553+113.  From left to right: Year of MAGIC observations; Effective time in hours; Integral flux above 150\,GeV in units of cm$^{-2}$ s$^{-1}$ and Crab Nebula \%, normalization factor f$_0$ in units of  cm$^{-2}$ s$^{-1}$ TeV$^{-1}$; in the last column, the $\Gamma$ index obtained by fitting the observed differential spectrum with a power law. The errors reported are statistical only. The systematic uncertainty is estimated to be $35\%$ in the flux level and 0.2 in the power index.
}  \label{table_int_spectra}
\vspace{0.15cm}
  \end{table*}

\section{VHE $\gamma$ ray results}
The 28.7~hours of good quality observations
of PG~1553+113 carried out between 2007 and 2009 resulted in a signal
of 8.8\,$\sigma$ of significance according to eq.\,17 of \citet{lima83},
obtained by combining the results from
each year, listed in Table~\ref{table_signal}.
The signal was extracted by analyzing the distribution of the parameter
\emph{alpha}, related to the incoming direction of the primary cosmic ray
inducing the atmospheric shower. 
More details on the signal extraction with the
\emph{alpha} technique can be found in \citealt{albert08b}.
For the signal detection, no cut in energy was applied.

The significance of the signal was 5.8\,$\sigma$ in 2007, 8.1\,$\sigma$ in 2008
and 4.2\,$\sigma$ in 2009. 
Due to a large difference in the energy thresholds
and changes in the experimental conditions, 
the obtained fluxes cannot be compared directly. 
A detailed spectral analysis is necessary in order to study
the source emission.

\subsection{Integral Flux}
\begin{figure*}[thb]
  \centering
  \includegraphics[width=7.0in]{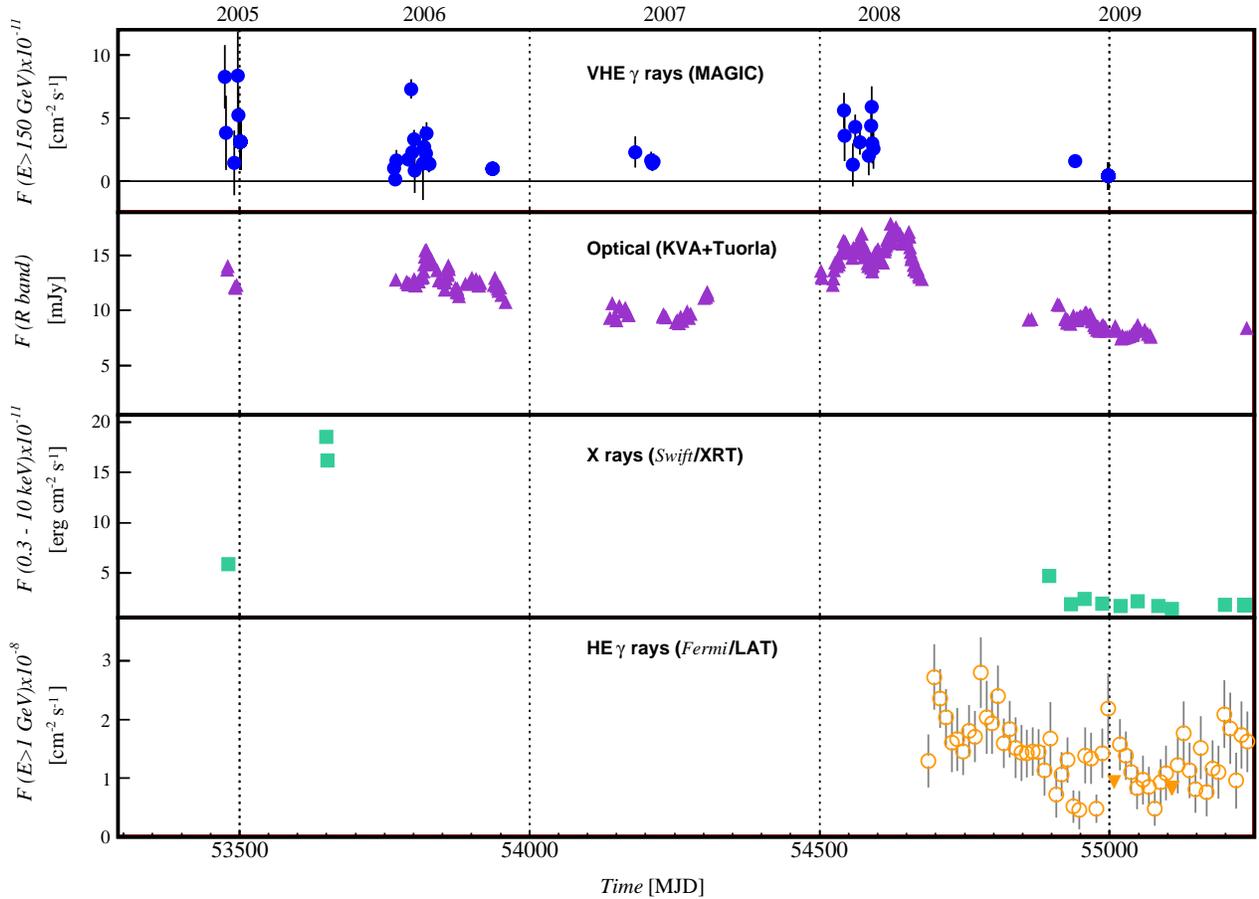}
  \caption{Multiwavelength light curve of PG~1553+113 from 2005 to late 2009. From upper to lower panel: VHE $\gamma$ rays above 150\,GeV measured by MAGIC (filled circles), optical flux in the R-band (triangles),   X rays 0.3--10\,keV flux  (squares) and $\gamma$ rays above 1\,GeV (open circles).  The error bars reported have 1\,$\sigma$ significance. Downward triangles, in lower panel, refer to 2\,sigma upper limits on the source flux above 1\,GeV.}
  \label{PG1553_mwl_LC}
 \end{figure*}

\begin{figure}[bht]
  \centering
  \includegraphics[width=3.5in]{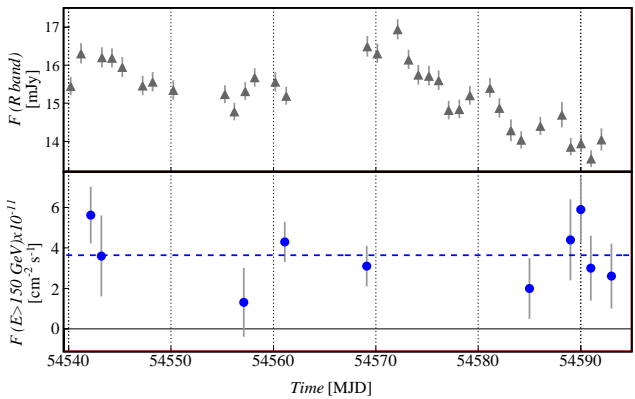}
  \caption{Zoom of Figure~\ref{PG1553_mwl_LC} around 2008 MAGIC observations (lower panel). In the upper panel the corresponding optical flux  has been superimposed for comparison.}
  \label{PG1553_LC_2008_zoom}
 \end{figure}
In order to explore the VHE $\gamma$ ray emission of PG~1553+113 
from each year, we compared the integral flux above 150\,GeV.
This value is a safe compromise taking into account the different
energy thresholds.
The final samples and the results of the spectral analyses are 
shown in Table~\ref{table_int_spectra}.

The integral fluxes measured above 150\,GeV 
lie in the range of 4\% to 11\% of the Crab Nebula flux measured by 
MAGIC \citep{albert08a}: the highest flux level is recorded in 2008 
(0.11 Crab units\footnote{The Crab unit used in this work 
  is an arbitrary unit obtained by dividing the integral energy flux measured
  above a certain threshold by the Crab Nebula flux 
  measured above the same threshold by MAGIC \citep{albert08a}.}), 
a factor between  two to three larger compared 
to the one measured in 2007 (0.04 Crab units)
and  2009 (0.05 Crab units). A constant fit to the data
has a probability smaller than 0.001\%.
Such changes in the flux level observed in  PG~1553+113
are quite moderate in comparison to other monitored
TeV blazars. For example, in Mkn~421 a
flux variations exceeding one order of magnitude
have been observed \citep{fossati08}.

A detailed study about possible flux level variations
on short timescale was carried out with the limiting
condition that the signal is not strong enough to allow for a 
detailed sampling on sub-day timescale.
The upper panel of Figure~\ref{PG1553_mwl_LC} displays
the light curve of PG~1553+113 measured from 2007 to 2009 by MAGIC with a
variable binning. For comparison, the daily flux levels measured in 
2005 and 2006 are shown, as 
extrapolated from the published data \citep{albert07},
and rescaled according to the power laws
that interpolate the differential fluxes. 
Furthermore, the 2006 mean integral flux above 150\,GeV 
taken during the multiwavelength campaign and
reported in \citet{albert09} is shown.
The former data have not been used for the integral flux study,
due to very large uncertainties
related to the extrapolation procedure.
We set 2-days, daily and monthly binning
for the 2007, 2008, and 2009 data sets respectively,
according to the significance of the signal.
The  2008 data are consistent with the hypothesis of constant flux with a probability of 50\% (Figure~\ref{PG1553_LC_2008_zoom}, lower panel).

\begin{figure*}[t!]
  \centering
  \includegraphics[width=3.5in]{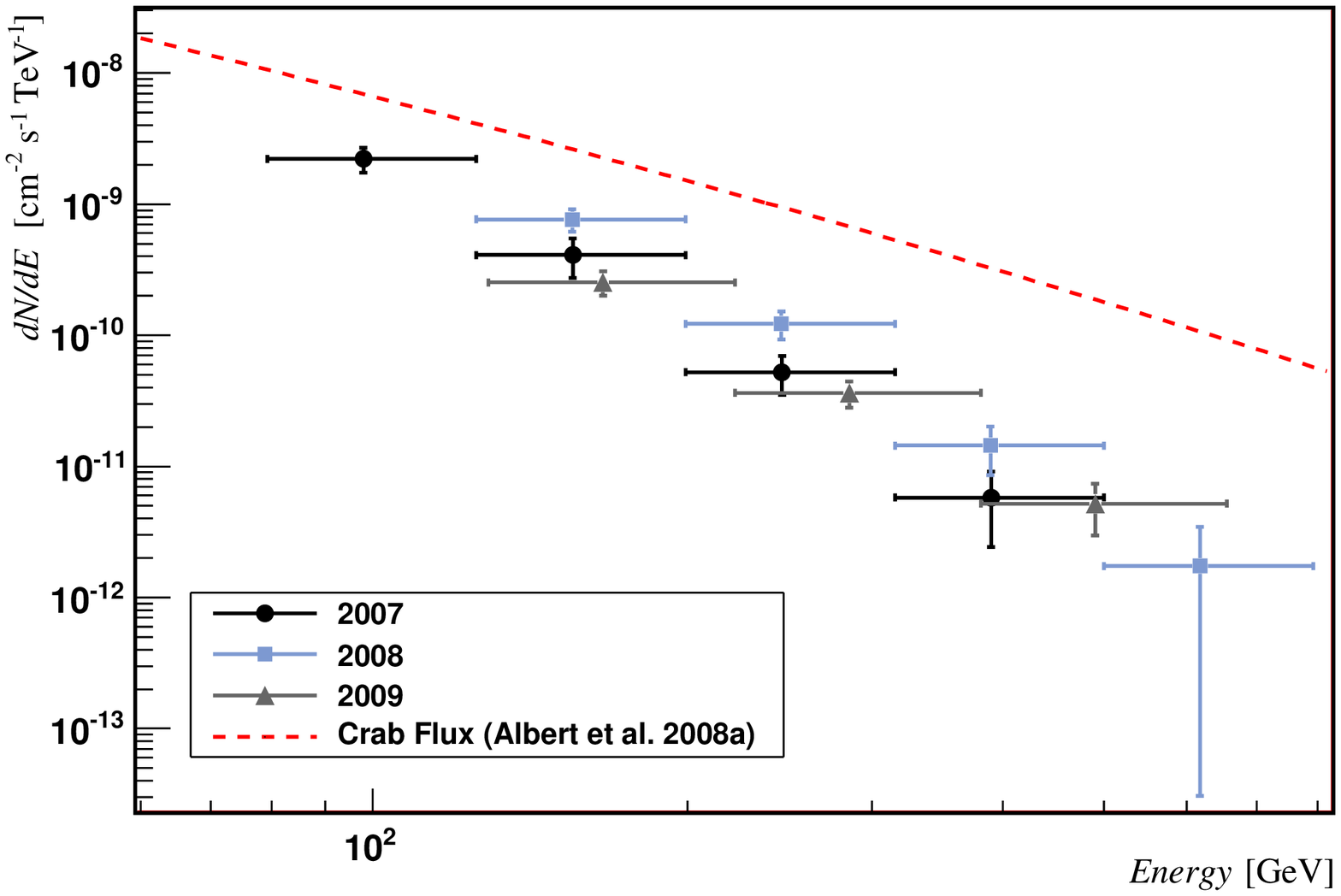}
  \includegraphics[width=3.5in]{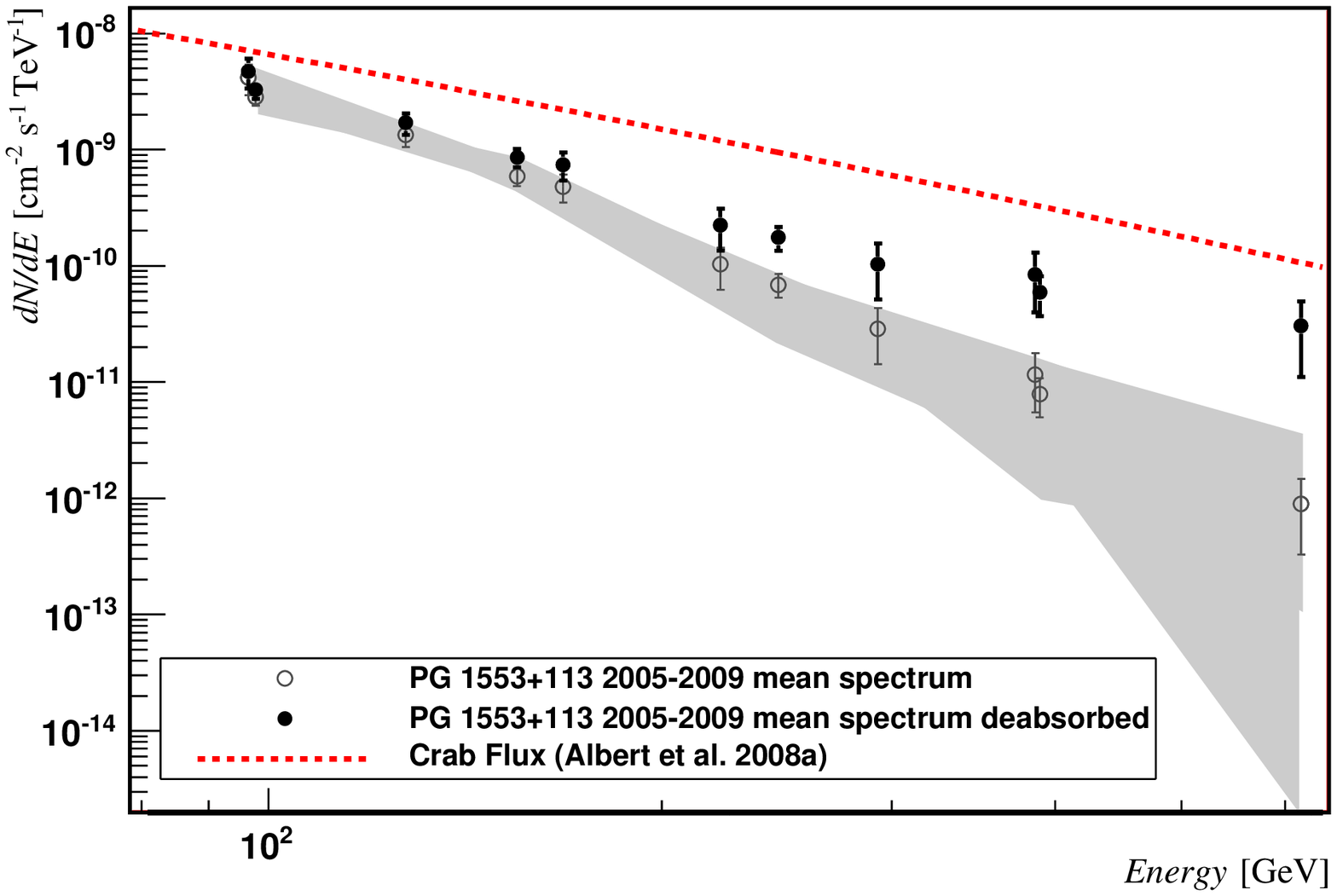}
  \caption{Differential energy spectra from PG~1553+113. Left Figure: comparison between 2007, 2008 and 2009 spectra. Right Figure: superimposition of 2005-2006 spectrum, from \citet{albert07}, to 2007-2009 mean spectrum and corresponding deabsorption for {\it z} = 0.4 using the  EBL model of \citet{dominguez10}.
 In both figures, the fit of the Crab Nebula spectrum measured by 
 MAGIC \citep{albert08a} is superimposed for comparison.}
  \label{PG1553_5years_spectra}
 \end{figure*}

In 2007, the observed time series is consistent with constant flux (93\%
of probability). 
Nothing similar can be concluded for the 2009 observations since
the significance of the signal is too low.

In general, the high energy threshold of the analysis together with
the weakness of the PG~1553+113 signal and its very steep spectrum 
make any variability study at short timescale difficult and 
might have hidden the detection of an increased activity 
on very short timescale.  

\subsection{Differential Flux}
The differential spectra observed from PG~1553+113 by MAGIC each
year from 2007 to 2009 are shown 
in the left plot of Figure~\ref{PG1553_5years_spectra}.

As for other blazars, each spectrum can be well fitted with a power law function of the form
 \begin{equation}
 \frac{dF}{dE}= f_0 * {\left( \frac{E}{200\,{\rm GeV}}\right)}^{-\Gamma}
 \label{simpeq}
\end{equation}
where $f_0$ is the flux at 200\,GeV and $\Gamma$ is the
power law index. The resulting indices are listed in the last
column of Table~\ref{table_int_spectra}.
The systematic uncertainty is estimated to be $35\%$ in the
flux level and 0.2 in the power index \citep{albert08a}, and is  
the sum of many  contributions, mainly related to the
use of MC simulations instead of test beams. 
Thanks to the low energy threshold of the analysis of
2007 data, the corresponding spectrum has a measured point below 100\,GeV. 
In particular, the measured differential energy flux at 
98\,GeV is (2.7~$\pm$~0.3)\,$\cdot$10$^{-9}$\,cm$^{-2}$s$^{-1}$TeV$^{-1}$, 
in agreement,  within the errors, with the 
low energy point measured in 2005 and 2006, 
(4.1~$\pm$~1.2)\,$\cdot$\,10$^{-9}$\,cm$^{-2}$s$^{-1}$TeV$^{-1}$ at 97 GeV.  

The 2008 differential energy spectrum 
measured above 150\,GeV has a slope of 4.3\,$\pm$\,0.4, while
the slope of the spectrum determined with a 
partially simultaneous sample taken during 
a multiwavelength campaign with other instruments 
is 3.4\,$\pm$\,0.1 between 70 to 350\,GeV \citep{aleksic10}.
The different energy range characterizing the measurements
fully accounts for this apparent disagreement: the 
spectral points measured in the range 150--350\,GeV are, in fact,
in very good agreement.

Finally, the 2009 differential spectrum is barely determined due to the
limited signal.  Except for the latter sample, whose
significance is rather low and
corresponding errors noticeably large, the power law indices
describing the spectra  are compatible. 
This indicates that the shape of the emitted spectrum does not
change, even if the total flux shows hints of (small amplitude) variability, as  
also noted for other BL Lacs such as 1ES~1218+304 \citep{acciari10}.

\begin{figure*}[bht]
  \centering
  \includegraphics[width=7.0in]{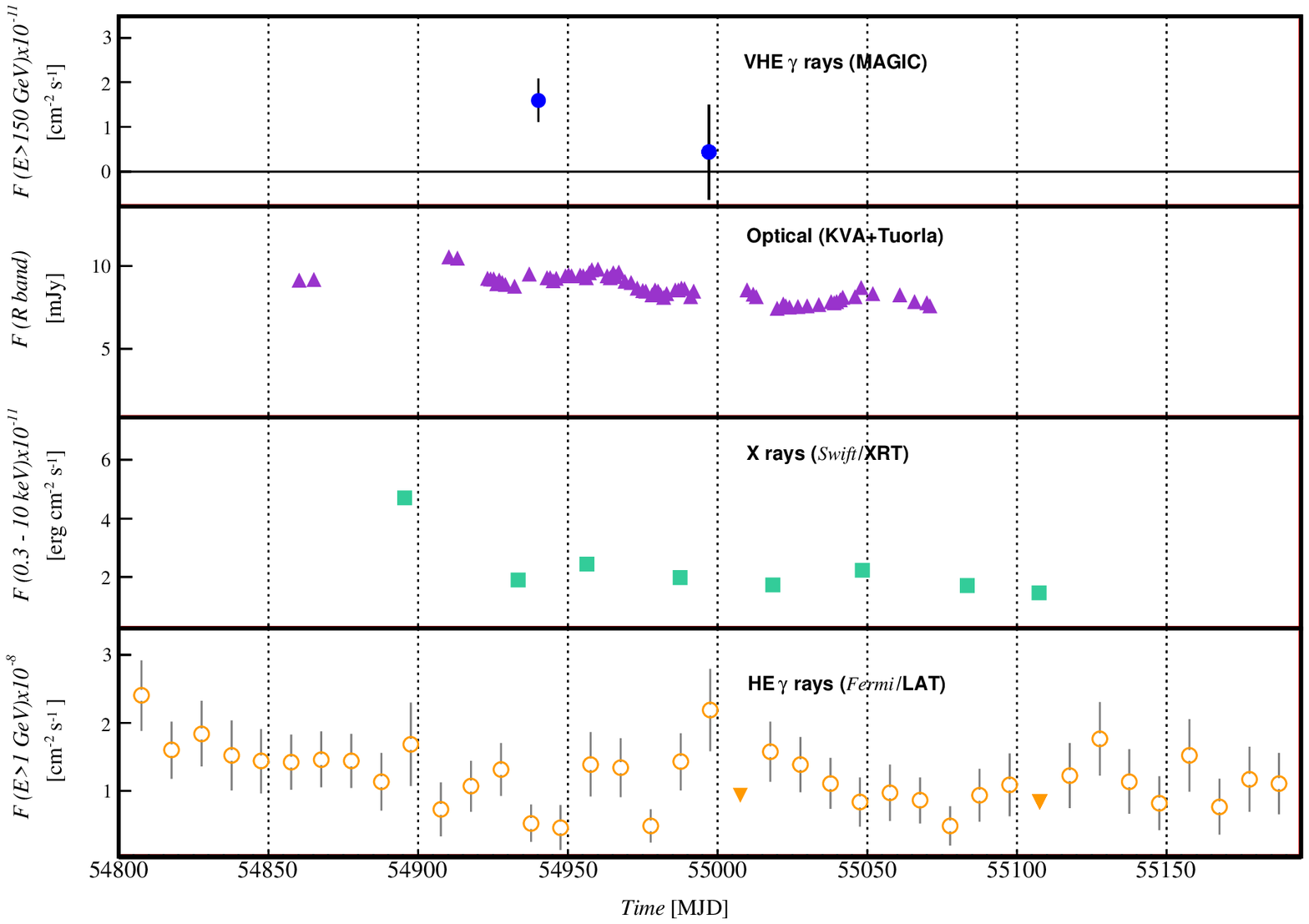}
  \caption{Zoom of Figure~\ref{PG1553_mwl_LC} around the Cycle V observations carried out in 2009. From upper to lower panel: VHE $\gamma$ rays above 150\,GeV measured by MAGIC (filled circles), optical flux in the R-band (triangles), X rays 0.3--10\,keV flux  (squares) and soft $\gamma$ rays above 1\,GeV (open circles).  Downward triangles, in lower panel, refer to 2\,sigma upper limits on the source flux above 1\,GeV.}
  \label{PG1553_mwl_LC_zoom}
 \end{figure*}

The right plot of Figure~\ref{PG1553_5years_spectra} 
shows the combined differential spectrum of PG~1553+113 from 2007 to 2009,
superimposed to the 2005-2006 spectrum
measured by MAGIC ($\Gamma$~=~4.21~$\pm$~0.25, \citealt{albert07}).
The gray band represents the systematic effect on the combined spectrum
resulting from the use
of different methods to correct for the effects due to the finite
energy resolution (procedure called {\it unfolding}, \citet{albert07b}).
The good agreement among these mean determinations suggests that despite the
(small) variability seen on yearly scale, the mean flux emitted by this source is stable.
A power law fit gives the values $\Gamma$~=~4.27~$\pm$~0.14 for the index and 
f$_0$~=~(1.61~$\pm$~0.14)~$\cdot$~10$^{-10}$~s$^{-1}$~cm$^{-2}$~TeV$^{-1}$ 
for the normalization factor, with a $\chi^2 /$dof = 5.7$/$9 and corresponding  
probability of 77\%. The integral flux above 150\,GeV is at the level of 8\% of the Crab Nebula flux. 

VHE photons from cosmological distances are absorbed 
in the interaction with the EBL. 
Taking into account the EBL absorption assuming the background
model proposed in \citet{dominguez10} and a redshift {\it z}~=~0.40, 
the intrinsic spectrum is compatible with a power law of index 3.09~$\pm$~0.20, 
as drawn in Figure~\ref{PG1553_5years_spectra}.
If we assume {\it z}~=~0.45, the corresponding spectrum 
is compatible with a power law of index 2.91~$\pm$~0.21.

\section{PG~1553+113 as seen at other wavelengths}

Figure~\ref{PG1553_mwl_LC} displays the light curve of PG~1553+113 in different wavelengths. The MAGIC data shown cover five cycles of TeV observations at energies above 150\,GeV.  The time bins used are variable, as described in the previous section.

The simultaneous optical R-band data are outlined in the second panel. These data are collected on a nightly basis by the Tuorla Observatory Blazar Monitoring Program\footnote{More information at http://users.utu.fi/kani/1m/} \citep{takalo07} using the KVA 35\,cm telescope at La Palma and the Tuorla 1\,meter telescope in Finland. 

For the third panel in Fig~1 we used the 
14 {\it Swift} pointed observations (Gehrels et al. 2004) of PG~1553$+$113
performed from 2005-04-20 to 2010-02-05. We summed the data collected on
5 and 7 July 2009 in order to have enough statistics to obtain a
good spectral fit. The XRT data were processed with standard procedures
(\texttt{xrtpipeline v0.12.6}), filtering, and screening criteria by using the
\texttt{Heasoft} package (v6.11). We consider the data collected in photon
counting mode, and thus only XRT event grades 0--12 were selected.

Source events were extracted from a circular region with a radius of 20 pixels
(1 pixel $\sim$ 2.36''), while background events were extracted from a circular
region with radius of 50 pixels away from the source region. 
Some observations showed an average count rate of  $>$ 0.5 counts s$^{-1}$,
thus pile-up correction was required. In that case we extracted the source
events from an annular region with an inner radius from 2 to 7 pixels
(depending on the source count rate and estimated by means of the PSF fitting
technique, see Moretti et al.~2005) and an outer radius of 30
pixels. We extracted background events within an annular region centered on
the source with radii 70 and 120 pixels. 
Ancillary response files were generated with \texttt{xrtmkarf}, and account for different extraction regions, vignetting and PSF corrections. We used the last spectral redistribution matrices in the Calibration database maintained by HEASARC. 
All spectra were rebinned with a minimum of 20 counts per energy bin to allow
$\chi^2$ fitting within XSPEC (v12.7.0; Arnaud 1996). 
We fit the spectrum with an absorbed (model \texttt{tbabs} in Xspec) log
parabola law (see e.g.~Tramacere et al.~2007), with a neutral hydrogen 
column fixed to its Galactic value (3.65$\times$10$^{20}$ cm$^{-2}$; Kalberla et
al.~2005). The observed 0.3-10 keV fluxes obtained by {\it Swift}/XRT in the
different observations are reported in the third panel of Fig.~1.

In the lower panel, the {\it Fermi}/LAT light curve of PG~1553+113, 
computed in 10-day bins, is displayed. {\it Fermi} data presented in this 
paper are restricted to the 1\,GeV\,-\,100\,GeV energy range and were 
collected from MJD 54682 (2008 August 4) to MJD 55200 (2010 January 4) 
in survey mode.

An unbinned analysis was performed to produce the light curve 
with the standard analysis tool {\it gtlike}, 
included in the Science Tools software package (version v09r21p00). 
P6\_V11 DIFFUSE  Instrument Response Functions (IRFs) were used, 
which are a refinement to previous analyses reflecting improved understanding of 
the point spread function and effective area (Abdo et al. 2011, in preparation).
For this analysis, only photons belonging to the Diffuse class and located in 
a circular Region Of Interest (ROI) of 10$^{\circ}$ radius, centered
at the position of PG 1553+113, were selected. In addition,
we excluded photons arriving from zenith angles $>$ 105$^{\circ}$
to limit contamination from Earth limb $\gamma$ rays, and photons with 
rocking angle $>$ 52$^{\circ}$ to avoid time intervals during which Earth entered the LAT FoV.

A separate analysis of the high energy emission in each time bin was performed. 
All point sources in the 1FGL within 15 degrees of PG~1553+113, 
including the source of interest itself,
were considered in the analysis. 
Those within the 10 degree radius ROI were fitted with a 
power law  with spectral indices
frozen to the values obtained from the likelihood analysis 
of the full data set,  while those beyond 10 degree radius
ROI had their values frozen to those found in 1FGL.

Upper limits at 2\,$\sigma$ confidence level 
(downward triangles in Figures 1 and 4) were 
computed for time bins with Test Statistics (TS)\footnote{
TS is 2 times the difference of the log(likelihood)
with and without the source \citep{mattox96}.} $<$~4, and were handled 
as in the first {\it Fermi}/LAT catalog paper. 
The estimated systematic uncertainty on the flux is 10\% at 100\,MeV, 
5\% at 500\,MeV, and 20\% at 10\,GeV.

As already noted, during the five years of monitoring  the source generally showed a marginal activity in the VHE $\gamma$ ray band. The same behavior is followed by the optical flux, whose variations are  limited within a factor of four, with a maximum flux reached in 2008 and a minimum value in 2009. A low emission in 2009 is also registered at all the other wavelengths, Figure~\ref{PG1553_mwl_LC_zoom}, suggesting that the source entered in a low activity state during that year, with a minimum reached few days after MAGIC observations. 

\begin{figure}[hbt]
  \centering
  \includegraphics[width=3.5in]{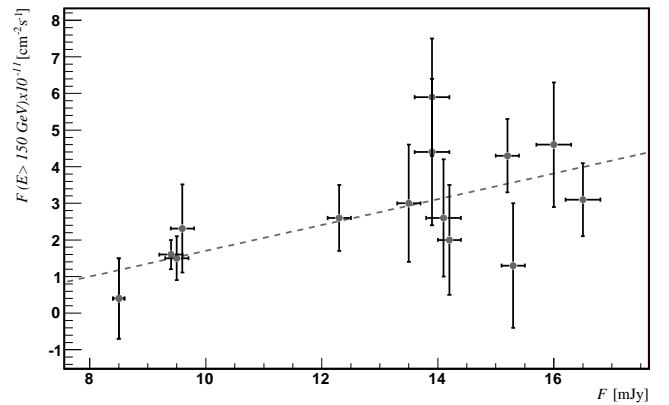}
  \caption{Correlation study between PG~1553+113 optical R--band flux and VHE $\gamma$ ray integral flux above 150\,GeV observed from 2006 to 2009.}
  \label{Correlation_OpticalTeV}
 \end{figure}

Figure~\ref{Correlation_OpticalTeV} shows the result of a correlation study
between optical and TeV simultaneous observations. 
The VHE $\gamma$ ray flux above 150\,GeV is plotted as a function of the optical flux. In order to increase statistics, we used for 2007/8/9 samples the daily light curve values; however, since the optical measurements have a different time coverage, in some cases we derived the mean VHE flux from two or more consecutive days. 2005 and 2006 data, from \citet{albert07}, were rejected from this study, due to the large uncertainty on the extrapolated flux in the VHE band. The mean flux value from 2006 multiwavelength campaign, reported in \citet{albert09}, is included. A linear relation among the two components has a 74\% probability,
which suggests a correlation between these two extreme energy bands. 
This result is in good agreement with the SSC model, which predicts a correlation between the synchrotron and the IC emission, related to the same electron population. Due to the poor simultaneity of VHE data with the other wavelengths, the same study has not been performed in X-rays and soft $\gamma$ rays. 

The X-ray light curve shows a pronounced variability, in contrast to optical and very high energy bands. The X-ray flux spans an interval of about one order of magnitude (with maximum in 2005 and minimum in 2009), larger than that observed in the TeV, optical and GeV bands. The different variability displayed by the synchrotron (X-ray) and inverse Compton (GeV-TeV) components seems to be somewhat in contrast with the typical behavior observed in TeV BL Lacs, showing, in general, a coordinated variability \citep{fossati08}\footnote{Rare exceptions to this rule are the so called ``orphan" TeV flares, e.g. \citet{kraw04}}. However, the sparse sampling of the observations and the lack of a truly simultaneous monitoring prevents any strong conclusion. In particular, no optical nor gamma-ray data are available during the period of the maximum X-ray flux in 2005. Coordinated multifrequency monitoring is necessary to further investigate this important issue.

In a dedicated paper \citep{abdo10b}, 
the \emph{Fermi}/LAT collaboration reported  
the analysis of the first year of PG~1553+113 data. They found that during the 
monitoring period the emission above 200\,MeV is almost steady. This is in 
contrast with the behavior of the source at higher energies ($>1$\,GeV). 
In our analysis of 2008 and 2009 LAT data (drawn in lower panel
of Figure~\ref{PG1553_mwl_LC}), in fact, a steady emission above 1\,GeV 
has a probability smaller than 0.1\% and is ruled out.
The lowest flux, observed in April 2009, has a value 
(0.5~$\pm$~0.3)~$\cdot$~10$^{-8}$~cm$^{-2}$s$^{-1}$, while 
the highest flux, detected in August 2008,
has a value (2.8~$\pm$~0.6)~$\cdot$~10$^{-8}$~cm$^{-2}$s$^{-1}$, more than 
5 times higher.
In our case, we are looking only at the upper edge of LAT band, probably 
close to the IC peak, while the integral flux above 200\,MeV reported by 
\citet{abdo10b} is dominated by lower energies, due to larger statistics. 
Therefore, we conclude that while at low energies (200\,MeV-1\,GeV) the IC continuum shows 
only marginal variability, this is not the case in the vicinity of the peak (some GeVs).
In fact, small variability at GeV energies is a common feature of HBLs 
\citep{abdo10a}.

\section{Modeling the SED}
\begin{figure*}[tbh]
  \centering
  \includegraphics[width=6.5in]{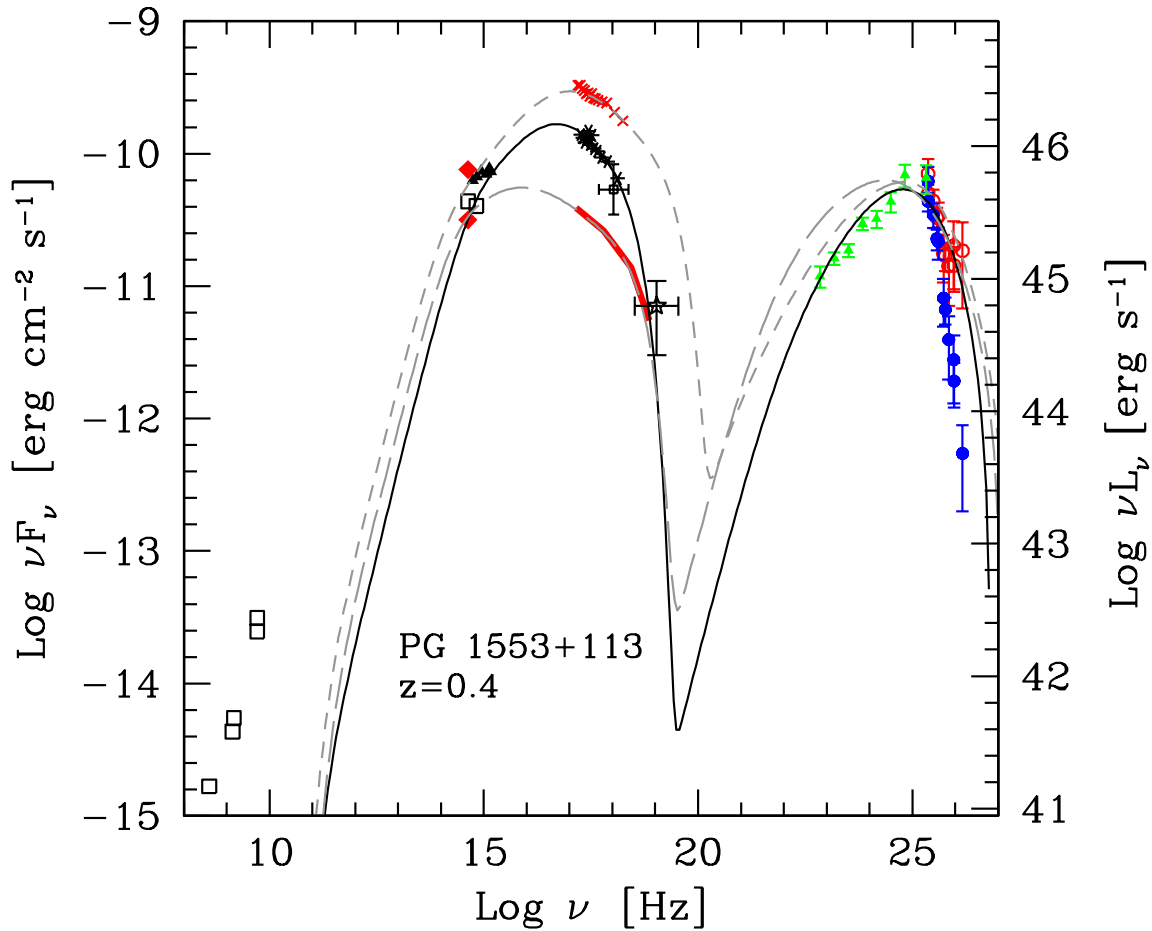}
  \vskip -2.0 truecm
  \caption{SED of PG 1553+113. Open black squares are radio-optical data from NED, red diamonds represent the KVA minimum and maximum flux measured in the period covered by MAGIC observations, together with optical-UV fluxes from {\it Swift}/UVOT (filled black triangles). For the X-ray data, two {\it Swift}/XRT spectra taken in 2005 (high flux state, red crosses, and intermediate state, black asterisks) are given, and a {\it Suzaku} spectrum taken in 2006 (continuous red line). The average 15--150 keV flux measured by {\it Swift}/BAT  (black star)  and the average {\it RXTE}/ASM flux between March 1 and May 31, 2008 (small black square), are also represented. The LAT spectrum averaged over $\sim$\,200 days (2008 August-2009 February) is given (green filled triangles). For MAGIC, we display the 2005-2006 and 2007-2009  observed spectra (blue filled circles) and the same spectra corrected for the absorption by the EBL (red open circles). The average SED is modeled with a one-zone SSC model (continuous black line). Alternative SED are superimposed in light gray. Detailed references are addressed in the text.} \label{PG1553_SED}
  \vskip 1.0 truecm
 \end{figure*}

In Figure~\ref{PG1553_SED}, we assembled the SED of PG~1553+113 using historical data and the MAGIC 
spectra described above. Open black squares displaying radio-optical 
data are from NED\footnote{http://nedwww.ipac.caltech.edu/}. 
In the optical band, we also show (red diamonds) the KVA 
minimum and maximum flux measured in the period covered by MAGIC 2005-2009
observations together with optical-UV fluxes from {\it Swift}/UVOT (filled black
triangles, from \citealt{tavecchio10}). For the X-ray data,  two 
{\it Swift}/XRT spectra taken in 2005 (high flux state, red crosses, and 
intermediate state, black asterisks, from \citealt{tavecchio10}) are given,
and a {\it Suzaku} spectrum taken in 2006 (continuous red line, 
from \citealt{reimer08}). 
In addition, the average 15--150 keV flux measured by {\it Swift}/BAT 
during the first 54 months of survey \citep{cusumano10} is shown (black star),
and the average {\it RXTE}/ASM flux between 
March 1 and May 31, 2008 (small black square), from quick--look results 
provided by the {\it RXTE}/ASM team\footnote{http://xte.mit.edu/asmlc/}.

The green triangles correspond to the LAT spectrum averaged over 
$\sim$\,200 days (2008 August-2009 February) from \citet{abdo10b}. As discussed in that paper, the
flux above 200 MeV is rather stable, showing very small variability over the entire period
of LAT observations.
It is likely that the variability observed at the highest energies 
is not important in determining the averaged 
spectrum due to the limited statistics.

For MAGIC, we report the  2005-2006 and 2007-2009 
observed spectra (filled circles) and the same spectra corrected for 
the absorption by the EBL using the model of \citet{dominguez10} (red open circles).

We model the SED with the one-zone SSC 
model fully described in \citet{maraschi03}. 
The emission zone is supposed to be spherical 
with radius $R$, in motion with bulk Lorentz factor $\Gamma$ at an angle 
$\theta $ with respect to the line of sight. Special relativistic effects 
are  described by the relativistic Doppler factor, $\delta=[\Gamma(1-\beta 
\cos \theta)]^{-1}$. The energy distribution of the relativistic 
emitting electrons is described by a smoothed broken power law function,
with limits $\gamma _{\rm min}$ and $\gamma _{\rm max}$ and break at $\gamma _{\rm b}$. 
To calculate the SSC emission, we use the full Klein--Nishina cross section. 

Given the large variations of the X-ray synchrotron flux, 
we decided to use the average
level of the synchrotron bump as measured by XRT, 
including also ASM and BAT fluxes 
to constrain the model. 
The corresponding input  parameters are listed in Table~\ref{param}. 
We also report the derived powers carried by the different components, 
relativistic electrons, $P_e$, magnetic field, $P_B$, and protons, $P_p$,
(assuming a composition of one cold proton per relativistic electron) 
and the total radiative luminosity $L_{\rm r}\simeq L_{\rm obs}/\delta^2$.

In order to investigate the role of different
parameters in the model, we have explored their variation
as a function of the intensity of the synchrotron peak.
To do so, we have modeled the SED
considering the two extreme states of the synchrotron
peak described above, respectively.
For the SSC peak, instead, we have fixed the VHE data.
The two curves representing the models
are superimposed in light gray to Figure~\ref{PG1553_SED}.
The parameters obtained, listed in the last two columns of
Table~\ref{param}, are quite similar to the ones obtained when considering the
average level of the synchrotron bump, except for the two variables
B and K, the magnetic field and the electron density.
Indeed, these two parameters regulate the relative importance of
 synchrotron and SSC
components. The state characterized by a low synchrotron emission has
larger B and smaller K values with respect to the mean state modeled
above. Conversely, the high synchrotron emission state
has smaller values of B and a larger K.

Finally, a comparison with the SED model obtained
in the multiwavelength campaign reported in \citet{aleksic10}, reveals that
the parameters used for building the two models are quite similar.
The major differences are the value of the Doppler factor, which
in our model is relatively higher ($\delta$\,$=$\,35) than in the previous one
($\delta$\,$=$\,23), and that of the magnetic field (0.5\,$G$ instead of 0.7\,$G$).
This difference is mainly due to the higher SSC peak frequency that we find in our data,
better defined by the combined LAT and MAGIC spectra.

 \begin{table}[thb]
 \centering
 \caption{Input model parameters for the models shown in Fig.~\ref{PG1553_SED}}
 \begin{tabular}{lcccc}
 \hline
 \hline
 Parameter & & Value$_{mean}$ & Value$_{max}$ & Value$_{min}$ \\
 \hline
 \hline
 $\gamma _{\rm min}$ &  $[10^3$]  &  $2.5$  & $1$ & $5$ \\
 $\gamma _{\rm b}$ & [$ 10^4$]    & $3.2$  & $3$ & $1.3$\\
 $\gamma _{\rm max}$ & [$ 10^5$]  & $2.2$  & $5.2$  & $4.1$ \\
 $n_1$ &                         & $2.0$  & $2.0$ & $2.0$  \\
 $n_2$ &                         & $4.0$  & $3.75$ & $3.55$ \\
 $B$ & [G]                       & $0.5$  & $0.8$ & $0.2$\\
 $K$ & [$ 10^3$ cm$^{-3}]$        & $5.35$ & $3.8$ & $25$ \\
 $R$ &  $[10^{16}$ cm]            & $1$  & $1$ & $1$ \\
 $\delta $ &                     & $35$   & $35$ & $35$ \\
 $P_{\rm e}$ &  [$10^{44}$ erg/s]  & $2.2$ &  &  \\
 $P_{B}$ &  [$10^{44}$ erg/s]     & $1.5$ &  &  \\
 $P_{\rm p}$ &  [$10^{44}$ erg/s]  & $0.34$  &  & \\ 
 $L_{\rm r}$ & [$10^{44}$ erg/s]   &  $6.3$ &  &  \\
 \hline
 \hline
 \end{tabular} 
 \vspace{0.10cm}
\footnotetext{We list for the three different models plotted in Fig.~\ref{PG1553_SED} the minimum, break and maximum Lorentz factors and the low and high energy slope of the electron energy distribution, the magnetic field intensity, the electron density, the radius of the emitting region and its Doppler factor. For the average model we also give the derived power carried by electrons, magnetic field, protons (assuming one cold proton per emitting relativistic electron) and the total radiative luminosity.} \label{param}
  \vspace{0.16cm}
\end{table}

The derived value of the total jet power, $P_{\rm jet}=P_{\rm e}+P_{B}+P_{\rm p}=4\times 10^{44}$ 
erg/s, is consistent with the typical values inferred modelling similar sources \citep{ghisellini10}. 
We use a relatively large minimum electron Lorentz factor $\gamma _{\rm min}\sim$~10$^3$ in order to 
reproduce the hard MeV-GeV continuum tracked by LAT (photon index $\Gamma~=~1.68~\pm~0.03$). 
The high value of $\gamma _{\rm min}$ implies that, as commonly derived in TeV BL Lacs, the relativistic electrons (and
the magnetic field, almost in equipartition) carry more power than the cold proton component.
Another characteristic that PG~1553+113 shares with the other TeV BL Lacs is that the total luminosity $L_{\rm r}$ is larger than 
the power supplied by electrons, magnetic field and protons. As discussed in \citet{celotti08}, this implies that 
either only a small fraction of leptons is accelerated at relativistic energies (leaving a reservoir of cold pairs and/or protons) or
that the jet is dissipating a large fraction of its power as radiation, eventually leading to the deceleration of the flow, as in fact observed 
at VLBI scales \citep{piner10} and envisaged in the models of structured jets \citep{georg03,ghisellini05}.

\section{Conclusions}
\vspace{0.16cm}
In this paper, we presented the analysis of three years of VHE $\gamma$ ray 
data of PG~1553+113 collected by MAGIC from 2007 to 2009. 
The data set was divided into individual years, and a significant signal 
was found in every sample, confirming PG~1553+113 as a stable presence in
the VHE sky.
The overall flux above 150\,GeV from 2007 to 2009
shows only a modest variability on yearly time-scale, within a factor 3, 
corresponding to a variation between 4\% to 11\% of the Crab Nebula flux. 
No clear variability on smaller time scales is evident in the sample.

For the spectral analysis, 
the data set was combined with previous observations carried out by MAGIC
during the first two Cycles of operations, from 2005 to 2006, for a total 
of five years of monitoring. This sample
was excluded from the temporal study due to very large systematics 
related to the flux extrapolation procedure.
Despite the hints of variability on the flux level, 
the differential flux from each year is in very good agreement with a 
power law of constant index 4.27~$\pm$~0.14. This behavior has been already
observed in other blazars, such as the HBL 1ES~1218+304 \citep{acciari10}.

PG 1553+113 was also monitored in optical, X-ray and soft $\gamma$ ray
frequencies, but only the former data could be used for correlation
studies thanks to the large timing coverage. Interestingly, a hint of
correlation with probability of 74\% was found between MAGIC and
R-band optical flux levels, which in turn shows only a modest
variability within a factor 4. A clear variability is seen in the
X-rays and $\gamma$ rays  above 1\,GeV.
The latter outcome, exploring the energies close to the IC peak, is only apparently in 
contradiction with previous results stating a quite stable spectrum for 
this source in the soft ($>200$ MeV) $\gamma$--ray band \citep{abdo10b}. 
The different energy thresholds used in the two studies can, in fact, 
explain very well the discrepancy, as discussed in the paper.

Finally, for the study of the spectral energy distribution, the mean differential spectrum  measured by MAGIC was combined with historical data at other wavelengths. Due to the large variations observed in X-rays and characterizing the synchrotron peak, we decided to use for the SED modeling the high energy bump, and the average level of the low energy bump. A more precise model requires coupling the VHE $\gamma$ ray part of the spectrum with  simultaneous  coverage of the synchrotron peak, in particular at optical-X-ray energies. An interesting feature of PG~1553+113 is the narrowness of the SSC peak derived from the LAT and MAGIC spectra, implying a relatively large value of the minimum Lorentz factor of the emitting electrons, $2.5\times 10^3$. This is also required by other HBLs with hard GeV spectra (e.g. Tavecchio et al. 2010). 

The MAGIC stereo system, with its increased sensitivity and low energy threshold, 
is the suitable instrument to further investigate eventual daily scale TeV 
variability, as well as to provide a good differential spectrum determination below 100\,GeV.

\section*{Acknowledgements}
\vspace{0.16cm}

We would like to thank the Instituto de Astrof\'{\i}sica de
Canarias for the excellent working conditions at the
Observatorio del Roque de los Muchachos in La Palma.
The support of the German BMBF and MPG, the Italian INFN, 
the Swiss National Fund SNF, and the Spanish MICINN is 
gratefully acknowledged. This work was also supported by 
the Marie Curie program, by the CPAN CSD2007-00042 and MultiDark
CSD2009-00064 projects of the Spanish Consolider-Ingenio 2010
programme, by grant DO02-353 of the Bulgaria
n NSF, by grant 127740 of 
the Academy of Finland, by the YIP of the Helmholtz Gemeinschaft, 
by the DFG Cluster of Excellence ``Origin and Structure of the 
Universe'', by the DFG Collaborative Research Centers SFB823/C4 and SFB876/C3,
and by the Polish MNiSzW grant 745/N-HESS-MAGIC/2010/0.

The \textit{Fermi} LAT Collaboration acknowledges generous ongoing support
from a number of agencies and institutes that have supported both the
development and the operation of the LAT as well as scientific data analysis.
These include the National Aeronautics and Space Administration and the
Department of Energy in the United States, the Commissariat \`a l'Energie Atomique
and the Centre National de la Recherche Scientifique / Institut National de Physique
Nucl\'eaire et de Physique des Particules in France, the Agenzia Spaziale Italiana
and the Istituto Nazionale di Fisica Nucleare in Italy, the Ministry of Education,
Culture, Sports, Science and Technology (MEXT), High Energy Accelerator Research
Organization (KEK) and Japan Aerospace Exploration Agency (JAXA) in Japan, and
the K.~A.~Wallenberg Foundation, the Swedish Research Council and the
Swedish National Space Board in Sweden.

Additional support for science analysis during the operations phase is gratefully
acknowledged from the Istituto Nazionale di Astrofisica in Italy and the Centre National d'\'Etudes Spatiales in France.

\end{document}